\documentclass[twocolumn,letter]{jpsj3} 

\title{Determination of the Upper Critical Field of a Single Crystal LiFeAs: The Magnetic Torque Study up to 35 Tesla}

\author{Nobuyuki KURITA$^{1,2}$\thanks{E-mail address: KURITA.Nobuyuki@nims.go.jp}, 
Kentaro KITAGAWA$^{2,3}$, Kazuyuki MATSUBAYASHI$^{2,3}$, Ade KISMARAHARDJA$^{4}$, Eun-Sang CHOI$^4$, James S. BROOKS$^4$, Yoshiya UWATOKO$^{2,3}$, Shinya UJI$^{1,2,5}$, Taichi TERASHIMA$^{1,2}$}

\inst{$^1$National Institute for Materials Science, Tsukuba, Ibaraki 305-0003, Japan \\
$^2$JST, Transformative Research-Project on Iron Pnictides (TRIP), Chiyoda, Tokyo 102-0075, Japan \\
$^3$Institute for Solid State Physics, The University of Tokyo, Kashiwanoha, Kashiwa, Chiba 277-8581, Japan \\
$^4$National High Magnetic Field Laboratory, Florida State University, Tallahassee, FL 32310, USA \\
$^5$Graduate School of Pure and Applied Sciences, University of Tsukuba, Ibaraki 305-0003, Japan}

\abst{We report on the upper critical field $B_\mathrm{c2}$ of a superconducting LiFeAs single crystal with $T_\mathrm{c}$\,$\sim$16\,K, determined from  magnetic torque measurements in dc-magnetic fields up to 35\,T and at temperatures down to 0.3\,K. 
$B_\mathrm{c2}$ at 0.3\,K is obtained to be 26.4\,T and 15.5\,T for the applied field $B_\mathrm{a}$\,$\parallel$\,$ab$ and $B_\mathrm{a}$\,$\parallel$\,$c$, respectively. 
The anisotropy parameter $\Gamma$\,=\,$B_\mathrm{c2}^\mathrm{ab}$/$B_\mathrm{c2}^\mathrm{c}$ is $\sim$\,3 at $T_\mathrm{c}$ and decreases to 1.7 as $T \rightarrow 0$, showing rather isotropic superconductivity.
While $B_\mathrm{c2}$ is orbitally-limited for $B_\mathrm{a}$ $\parallel$\,$c$, the spin-paramagnetic effect is evident in the temperature dependence of $B_\mathrm{c2}$ for $B_\mathrm{a}$\,$\parallel$\,$ab$.
}

\kword{iron-based superconductor, LiFeAs, upper critical field, anisotropy}

\begin{document}
\maketitle


Since the discovery of superconductivity in LaFeAs(O,F) with $T_\mathrm{c}$\,=\,26\,K\cite{Kamihara}, a variety of related compounds containing FeAs-layers has been found to exhibit superconductivity\cite{review}.
The parent compounds $R$FeAsO ($R$\,=\,rare earth, ``1111" system) and $A$Fe$_2$As$_2$ ($A$\,=\,alkaline earth or Eu, ``122" system) with the ZrCuSiAs- and ThCr$_2$Si$_2$-type structures, respectively, undergo antiferromagnetic and structural transitions.
The transitions can be suppressed by several kinds of doping effects\cite{Kamihara, Sefat2008, Rotter2008b, Sasmal2008, Jiang_BaFe2AsP2} or application of pressure\cite{Alireza,
Matsubayashi, Terashima}, and $T_\mathrm{c}$ reaches $\sim$\,56\,K in some compounds\cite{Kito2008, ZARen2008a,Wang_56K}.
The magnetic long range order usually competes with superconductivity, but the fluctuation likely plays a crucial role in the pairing mechanism of the Fe-based high-$T_\mathrm{c}$ systems.
It is also of interest that, around the optimal condition where $T_\mathrm{c}$ shows its maximum, deviation from conventional Fermi-liquid behavior has been observed such as $\rho$\,$\sim$\,$T$\cite{Sm1111_NFL,SrK122_NFL}, anomalous Hall angle\cite{Sm1111_NFL}, an enhancement of effective masses\cite{Shishido_dHvA} etc.
These superconducting and normal-state features bear resemblance with those widely reported for strongly-correlated electron systems including cuprates and heavy fermion compounds.

The title compound LiFeAs, categorized into the ``111" system with CeFeSi-type structure, has distinctive characteristics:
(i) the stoichiometric superconductivity with $T_\mathrm{c}$ as high as $\sim$\,17\,K\cite{Wang_SSC}, 
(ii) no experimental evidence for the magnetic/structural transitions\cite{Tapp_poly,Pratt_muon}, and
(iii) single crystals with high quality (residual resistivity ratio up to 50).\cite{Song_APL,imai;condmat}
Therefore, LiFeAs provides a unique opportunity to probe the intrinsic properties of the Fe-based high-$T_\mathrm{c}$ superconductivity.

Up to now,  there are few reports on systematic measurements of the upper critical field $B_\mathrm{c2}$ of Fe-based superconductors to address the issue of the pair-breaking mechanism. 
This is mainly due to the fact that high-$T_\mathrm{c}$ superconductors including Fe-based systems generally have extremely high $B_\mathrm{c2}$.
In many cases, accordingly, the low-temperature behavior is extrapolated from the high temperature data around $T_\mathrm{c}$, which may lead to misleading conclusions.
A precise determination of $B_\mathrm{c2}$ over a whole temperature range could provide important clues to the pair-breaking mechanism of high-$T_\mathrm{c}$ superconductivity.

Here, we present the first report on the whole temperature dependence of $B_\mathrm{c2}$ and its angular variation of a LiFeAs single crystal, using a magnetic torque technique with a 35\,T dc resistive magnet.

\begin{figure}
\begin{center}
\includegraphics[width=0.9\linewidth]{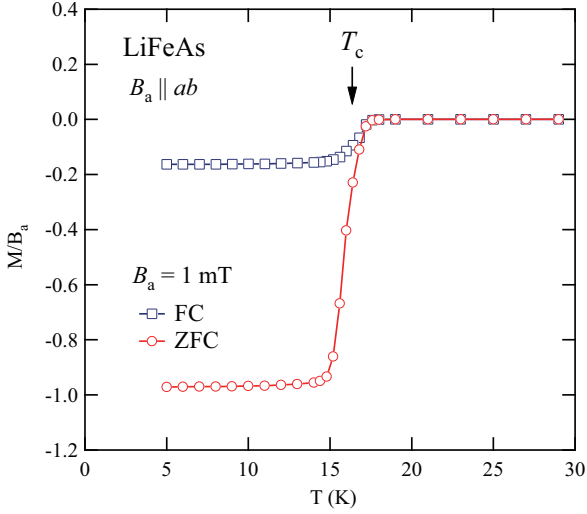}
\end{center}
\caption{(Color online) Magnetization divided by applied magnetic field, $M$/$B_\mathrm{a}$, vs $T$ of a single crystal LiFeAs around the transition temperature, $T_\mathrm{c}$, in a field of $B_\mathrm{a}$\,=\,1\,mT applied parallel to the $ab$ plane. 
Results measured in zero-field-cooled (ZFC) and field-cooled (FC) processes are displayed. 
A tiny piece cleaved from the same crystal was mounted on a microcantilever for the magnetic torque measurement as shown in the inset photograph.  
The field angle $\theta$ is the angle between the $c$ axis and the applied magnetic field $B_\mathrm{a}$.} \label{fig1}
\end{figure}

Single crystals LiFeAs were prepared from high-purity constituent elements in a ratio of Li:Fe:As\,=\,2:1:2  by a self-flux method.\cite{imai;condmat,Sample}.
The residual resistivity ratio is as high as 45, showing the high quality of the crystals.
Temperature dependence of the dc magnetic susceptibility of LiFeAs was measured with an applied field $B_\mathrm{a}$ of 1\,mT for $B_\mathrm{a}$\,$\parallel$\,$ab$, in a Magnetic Property Measurement System (MPMS: Quantum Design) around $T_\mathrm{c}$.
As displayed in the main panel of Fig.~\ref{fig1}, the bulk superconductivity was confirmed from the clear diamagnetic transition below $T_\mathrm{c}$\,$\sim$\,16\,K with approximately 16\,\% (97\,\%)  of the superconducting  Meissner (shielding) volume fraction for (zero-)field-cooled process.
A tiny piece taken from the same crystal was mounted on a piezoresistive microcantilever with a small amount of grease (see, inset photo of Fig.~\ref{fig1}), attached to a rotator probe, and inserted into a $^3$He cryostat (Heliox: Oxford).
The angle $\theta$ is the angle between the applied field $B_\mathrm{a}$ and the $c$ axis: $\theta$\,=\,0\,$^{\circ}$ for $B_\mathrm{a}$\,$\parallel$\,$c$ and $\theta$\,=\,90\,$^\circ$ for $B_\mathrm{a}$\,$\parallel$\,$ab$.
The magnetic torque was measured with a water-cooled dc resistive magnet in fields up to 35\,T and at temperatures down to 0.3 K.
Since the magnetic torque, proportional to $M$\,$\times$\,$B_\mathrm{a}$, is very small at $\theta$\,=\,0\,$^{\circ}$, the torque is measured at $\theta$\,=\,5\,$^{\circ}$ to improve the signal-noise ratio, which is specified as $B_\mathrm{a}$\,$\parallel$\,$c$. 
We confirmed that the difference of $B_\mathrm{c2}$ between the two angles is negligible from the angular dependence of $B_\mathrm{c2}$.


\begin{figure}
\begin{center}
\includegraphics[width=0.9\linewidth]{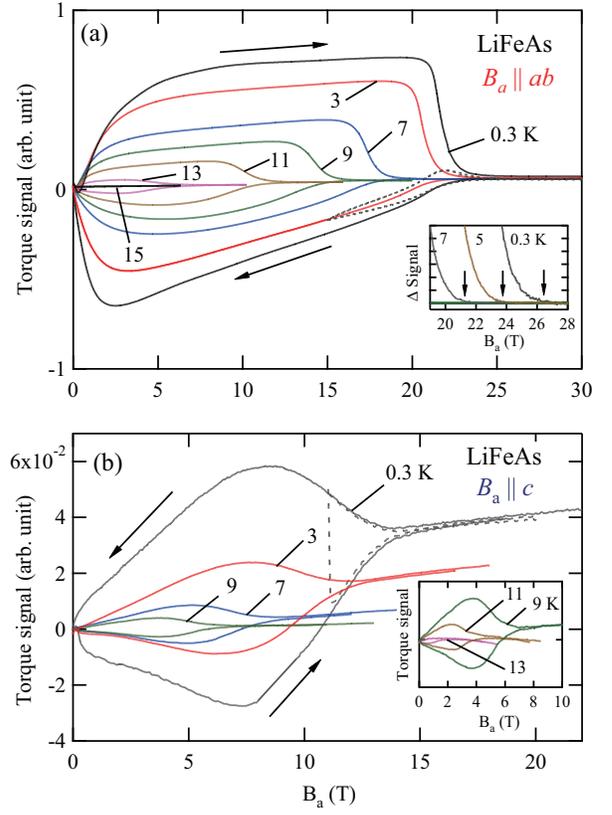}
\end{center}
\caption{(Color online) Magnetic torque signals of LiFeAs as a function of $B_\mathrm{a}$ at various temperatures down to 0.3\,K for (a) $B_\mathrm{a}$\,$\parallel$\,$ab$ and (b) $B_\mathrm{a}$\,$\parallel$\,$c$ (inset: magnified view). 
The solid curves represent the data obtained in field-up and -down sweeps between $B_\mathrm{a}$\,=\,0 and above $B_\mathrm{c2}$,
where the arrows indicate the sweep directions. 
Small-loop measurements around $B_\mathrm{c2}$, as shown by the dashed curves at $T$\,=\,0.3\,K, were performed at each temperature.
In this study, $B_\mathrm{c2}$ was defined as the field where the difference between field\,-\,up and -\,down signals in a small loop becomes zero within experimental error as illustrated in the inset of (a). 
} \label{fig2}
\end{figure}

Figure~\ref{fig2} shows the field dependence of the torque signals of LiFeAs up to 35\,T at fixed temperatures down to 0.3\,K for (a)\,$B_\mathrm{a}$\,$\parallel$\,$ab$ and (b)\,$B_\mathrm{a}$\,$\parallel$\,$c$.
One can see highly hysteretic behavior between the field\,-\,up and -\,down sweeps, which is more significant at low temperatures.\cite{TorqueSignal}
The hysteresis of the torque response, the irreversible curve, appears in the superconducting mixed state when the pinning force is strong enough to trap the flux lines.
In this study, we define the upper critical field $B_\mathrm{c2}$ as the field where the irreversibility disappears.\cite{Bc2}
To accurately determine $B_\mathrm{c2}$ by minimizing a drift of the torque signals, we made the small-loop measurements around $B_\mathrm{c2}$ at each temperature, as shown by the dashed curves in (a) and (b) for $T$\,=\,0.3\,K, in addition to the large-loop ones between $B_\mathrm{a}$\,=\,0 and above $B_\mathrm{c2}$. 
The definition of $B_\mathrm{c2}$ is illustrated in the inset of Fig.~\ref{fig2}(a), where the vertical axis represents the difference of torque signals between field-up and -down sweeps of a small loop.
Note that the shapes of the small loops near $B_\mathrm{c2}$ for $B_\mathrm{a}$\,$\parallel$\,$ab$ and \,$B_\mathrm{a}$\,$\parallel$\,$c$ are quite different.
It could be possible that the pinning mechanism is different for the two field orientations, although the detail remains to be clarified.

\begin{figure}
\begin{center}
\includegraphics[width=0.9\linewidth]{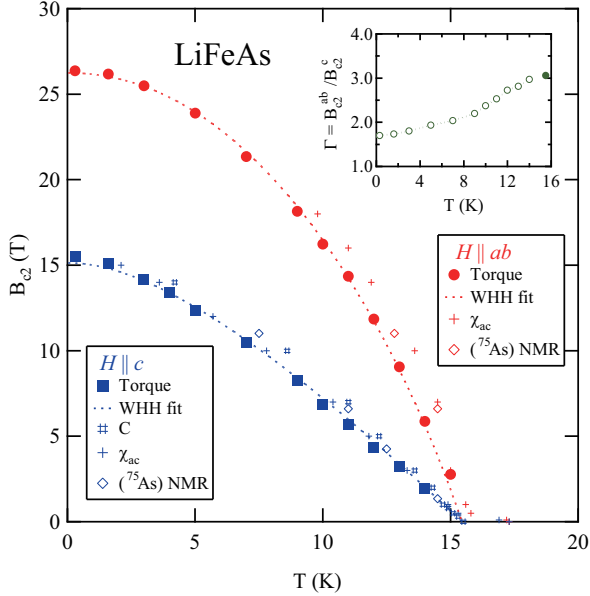}
\end{center}
\caption{(Color online)$B_\mathrm{c2}$ vs $T$ of LiFeAs for $B_\mathrm{a}$\,$\parallel$\,$ab$ and $B_\mathrm{a}$\,$\parallel$\,$c$. 
The dashed curves indicate fits to the data based on the WHH theory (see text). 
For comparison, data deduced from $C$, $\chi_\mathrm{ac}$, and $^{75}$As-NMR measurements are shown\cite{MK}. 
The inset shows the $T$-dependence of the anisotropy parameter $\Gamma$\,=\,$B_\mathrm{c2}^{ab}$/$B_\mathrm{c2}^{c}$. 
The data at $T_\mathrm{c}$ (\textbullet) is obtained from the WHH fit.} \label{fig3}
\end{figure}

Figure~\ref{fig3} displays thus determined $B_\mathrm{c2}$ of LiFeAs as a function of $T$ for $B_\mathrm{a}$\,$\parallel$\,$ab$ and $B_\mathrm{a}$\,$\parallel$\,$c$.
The $B_\mathrm{c2}$ curves are consistent with the results obtained by specific heat ($C$), ac-susceptibility ($\chi_\mathrm{ac}$), and $^{75}$As nuclear magnetic resonance (NMR) measurements using samples of the same batch\,\cite{MK}, and are qualitatively similar to other results\cite{Lee_LiFeAs,Heyer_LiFeAs}.
As indicated by the dashed curves, the data can be well fitted using a Werthamer-Helfand-Hohenberg
(WHH) formula containing the spin-paramagnetic and orbital pair-breaking effects\cite{WHH1966}.
The fits give the Maki parameter $\alpha$\,=\,2.30 and 0.75 for $B_\mathrm{a}$\,$\parallel$\,$ab$ and $B_\mathrm{a}$\,$\parallel$\,$c$, respectively, where $T_\mathrm{c}$\,=\,15.5\,K is fixed\cite{Tc}.
Because the torque signal, $\propto$\,$M$\,$\times$\,$B_\mathrm{a}$, diminishes at low fields, it is difficult to unambiguously determine the initial slope  d$B_\mathrm{c2}$/d$T|_{T=T_\mathrm{c}}$.
Instead, the values of d$B_\mathrm{c2}$/d$T|_{T=T_\mathrm{c}}$ are estimated to be 4.43 and 1.44\,T/K for $B_\mathrm{a}$\,$\parallel$\,$ab$ and $B_\mathrm{a}$\,$\parallel$\,$c$, respectively, from the relation $\alpha$\,=\,$-$0.52\,$\mathrm{d}B_\mathrm{c2}/\mathrm{d}T|_{T=T_\mathrm{c}}$.
These values are comparable to those obtained from ac-$\chi$.\cite{MK}
The orbital critical fields $B_\mathrm{c2}^*$ at $T$\,=\,0 are estimated to be 47.2 and 15.3\,T for $B_\mathrm{a}$\,$\parallel$\,$ab$ and $B_\mathrm{a}$\,$\parallel$\,$c$, respectively, from a dirty limit formula $B_\mathrm{c2}^*$(0)\,=\,0.69\,$T_\mathrm{c}$\,$\mathrm{d}H_\mathrm{c2}/\mathrm{d}T|_{T=T_\mathrm{c}}$.\cite{WHH1966}
The Pauli-Clogston paramagnetic limit $B_\mathrm{po}$\,=\,1.84\,$T_\mathrm{c}$\,\cite{CClimit1} is 28.5\,T.
The Ginzburg-Landau (GL) coherence length $\xi$ is obtained to be $\xi_{ab}$\,=\,4.64\,nm and  $\xi_c$\,=\,1.50\,nm, using $B_\mathrm{c2}^{*\,ab}(0)$\,=\,$\mathrm{\Phi}_0$/2$\pi\xi_{ab}(0)\xi_{c}$(0) and $B_\mathrm{c2}^{*\,c}(0)$\,=\,$\mathrm{\Phi}_0$/2$\pi\xi_{ab}(0)^2$, where $\mathrm{\Phi}_0$\,=\,2$\pi\hbar$/2$e$\,=\,\,2.07\,$\times$\,10$^{-15}$\,T\,m$^2$ is the flux quantum. 
At $T$\,=\,$T_\mathrm{c}$, the anisotropy of the effective masses, $m^*_{ab}$/$m^*_{c}$, is about 0.11, using $B_\mathrm{c2}^{*\,ab}$\,/\,$B_\mathrm{c2}^{*\,c}$\,=\,($m_{c}^*$/$m_{ab}^*$)$^{0.5}$.
These superconducting parameters are summarized in the Table I, together with those of a stoichiometric superconductor KFe$_2$As$_2$,\cite{K122} for comparison.

\begin{table}
\caption{Superconducting parameters of LiFeAs and KFe$_2$As$_2$\,\cite{K122}, for comparison, obtained from the WHH fits; $\alpha$: Maki parameter, $\lambda_\mathrm{so}$: Spin-orbit scattering parameter, $B_\mathrm{c2}^*$: Orbital critical field, $B_\mathrm{po}$: Paramagnetic critical field, $\xi$: GL coherence length.}
\label{t1}
\begin{tabular}{p{7em}ccp{1em}cc}
\hline
 &  \multicolumn{2}{c}{LiFeAs}   & & \multicolumn{2}{c}{KFe$_2$As$_2$\,\cite{K122}}  \\
 &  $B$\,$\parallel$\,$ab$  & $B$\,$\parallel$\,$c$& &$B$\,$\parallel$\,$ab$ &$B$\,$\parallel$\,$c$\\
\hline
$\alpha$ &   2.30     & 0.75  & &  2.30&0.340\\

$\lambda_\mathrm{so}$ &    0.51 &     $\infty$ &  &0.36 &$\infty$\\

$T_\mathrm{c}$\,(K) &    \multicolumn{2}{c}{15.5}  &  &\multicolumn{2}{c}{2.79} \\

$-$\,$\frac{dB_\mathrm{c2}}{dT}|_{T_\mathrm{c}}$\,(T/K)  &  4.43     & 1.44  & & 3.8 & 0.71 \\

$B_\mathrm{c2}$\,$(\mathrm{0.3\,K})$\,(T)  & 26.4    & 15.5  & &4.40  &1.25 \\

$B_\mathrm{c2}^*$(0)\,(T) & 47.2    & 15.3  && 8.44 & 1.25\\

$B_\mathrm{po}$\,(T) &   \multicolumn{2}{c}{28.5}    &&  \multicolumn{2}{c}{5.13}\\

$\xi$\,(nm) & 4.64     & 1.50  & & 16.3 &2.45\\
\hline
\end{tabular}     
\end{table}

It is interesting to note that $B_\mathrm{c2}^{*\,c}(0)$\,$<$\,$B_\mathrm{po}$\,$<$\,$B_\mathrm{c2}^{*\,ab}(0)$.
This relation results in the orbitally limited $B_\mathrm{c2}$ for $B_\mathrm{a}$\,$\parallel$\,$c$, as indicated by $\lambda_\mathrm{so}^c$\,=\,$\infty$, and strongly spin-paramagnetically limited $B_\mathrm{c2}$ for $B_\mathrm{a}$\,$\parallel$\,$ab$, as indicated by that $B_\mathrm{c2}^{ab}$(0.3\,K)\,$\ll$\,$B_\mathrm{c2}^{*\,ab}$(0),
and effectively reduces the $B_\mathrm{c2}$ anisotropy at low temperatures.
A similar trend, namely the trend that the weaker orbital effect for $B_\mathrm{a}$\,$\parallel$\,$ab$ is partly compensated for by the spin-paramagnetic effect to yield a reduced anisotropy, is clearly seen in a stoichiometric superconductor KFe$_2$As$_2$ with low $T_\mathrm{c}$ of 2.8\,K\,\cite{K122} (see, Table I), as well as in the high-$T_\mathrm{c}$ systems including (Ba,K)Fe$_2$As$_2$\cite{Yuan_BaK122,Altarawneh_BaK122}, Ba(Fe,Co)$_2$As$_2$\cite{BaFeCo2As2_Yamamoto,Kano},  and ``11"-type Fe(Se,Te)\cite{Fang_FeSe,Khim_FeSe,Kida_FeSe,Klein_FeSe}.

\begin{figure}
\begin{center}
\includegraphics[width=0.9\linewidth]{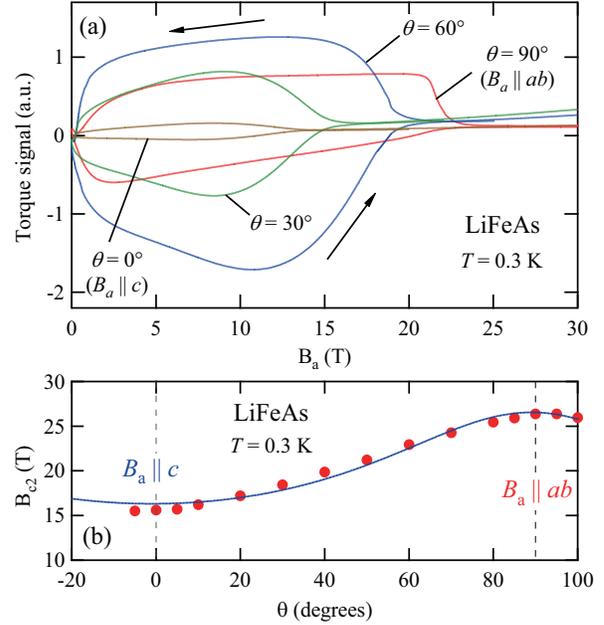}
\end{center}
\caption{(Color online)  (a) Torque signals of LiFeAs as a function of $B_\mathrm{a}$ at 0.3\,K  for several angles $\theta$, where $\theta$\,=\,0 and 90\,$^{\circ}$ correspond to $B_\mathrm{a}$\,$\parallel$\,$c$ and $B_\mathrm{a}$\,$\parallel$\,$ab$, respectively.
(b) $\theta$-dependence of $B_\mathrm{c2}$ determined at $T$\,=\,0.3\,K.  
The solid curve is a fit by a GL formula (see text). } \label{fig4}
\end{figure}

Figure~\ref{fig4} shows (a) the torque signals vs $B_\mathrm{a}$ at 0.3\,K for several angles, $\theta$, and (b) $B_\mathrm{c2}$ vs $\theta$ at 0.3\,K.
The solid curve in Fig.~\ref{fig4}(b) is a fit to the $B_\mathrm{c2}(\theta)$ data using a Ginzburg-Landau (GL) formula,\cite{Decroux1966}
$B_\mathrm{c2}(\theta)$\,=\,$B_\mathrm{c2}^\mathrm{c}$/(cos$^2$\,$\theta$\,+\,$\Gamma^{-2}$\,sin$^2$\,$\theta$)$^{0.5}$.
The fit yields $B_\mathrm{c2}^\mathrm{c}$\,=\,16.3\,T and $\Gamma$\,=\,1.63.
Note that the GL formula is based on the orbital effect whereas the spin-paramagnetic effect is crucial for $B_\mathrm{a}$\,$\parallel$\,$ab$ in the present case.
The quality of the fit is therefore rather limited as indicated by the deviations of the data points from the fit curve.
The inset of Fig.~\ref{fig3} shows the $T$-dependence of the anisotropy parameter $\Gamma$, defined as $\Gamma$\,=\,$B_\mathrm{c2}^{ab}$\,/\,$B_\mathrm{c2}^{c}$.
As temperature decreases, $\Gamma$ decreases from $\Gamma$\,=\,3.1 at $T_\mathrm{c}$ (from the WHH fit). 
At 0.3\,K, $B_\mathrm{c2}^{ab}$\,=\,26.4\,T and $B_\mathrm{c2}^{c}$\,=\,15.5\,T, which give $\Gamma$\,=\,1.7.
The low-temperature variation of $\Gamma$ is similar to those observed in the ``122" and ``11" systems\,\cite{Yuan_BaK122,Altarawneh_BaK122,BaFeCo2As2_Yamamoto,Kano,Fang_FeSe,Khim_FeSe,Kida_FeSe,Klein_FeSe},
but the value of  $\Gamma$ in LiFeAs is slightly larger.
In the ``122" and ``11" systems, $\Gamma$\,=\,2\,$\sim$\,3 at $T_\mathrm{c}$ and $\Gamma$ appears to approach $\sim$\,1 at 0\,K\,\cite{Yuan_BaK122,Altarawneh_BaK122,BaFeCo2As2_Yamamoto,Kano,Fang_FeSe,Khim_FeSe,Kida_FeSe,Klein_FeSe}.
This is markedly different from the results reported for the ``1111" system.
In NdFeAs(O,F), for example, $\Gamma$ is 9.2 at $T_\mathrm{c}$\,\cite{Jaroszynski_1111} (no report for $\Gamma$ as $T \rightarrow 0$ because of the large $B_\mathrm{c2}$).
The larger value of $\Gamma$ in the  ``1111" system can be ascribed to the more two-dimensional Fermi surface structures than those in the  ``111", ``122" and ``11" systems, as expected from band structure calculations\,\cite{Singh_LDA}.
Similarly, the slightly larger $\Gamma$ of LiFeAs than those of the  ``122" and ``11" systems might be due to the more two-dimensionality of LiFeAs\,\cite{Singh_LDA}.
In KFe$_2$As$_2$, however, there is an exceptionally large anisotropy ($\Gamma$\,=\,6.8 at $T_\mathrm{c}$, see Table I) for the ``122" system\,\cite{K122}, which is, on the other hand, in accord with the large resistivity anisotropy $\rho_{c}/\rho_{ab}$\,$\sim$\,40\,\cite{K122}.
In contrast, $\rho_{c}/\rho_{ab}$ is reported to be as small as 3.3 in LiFeAs, and 2\,$\sim$\,5 in Ba(Fe,Co)$_2$As$_2$\,\cite{Tanatar_PRB}.


To conclude, we have performed high-field magnetic torque measurements of a LiFeAs single crystal up to 35\,T, and determined the $T$- and $\theta$-dependence of $B_\mathrm{c2}$ down to 0.3\,K.
The anisotropy parameter $\Gamma$\,=\,$B_\mathrm{c2}^{ab}/B_\mathrm{c2}^{c}$ is slightly larger than typical values found in the ``122" and ``11" systems, but is still small, indicating quasi-isotropic superconductivity.
Its temperature dependence is similar to those observed in the ``122" and ``11" systems.
The detailed analyses show that, while $B_\mathrm{c2}$ for $B_\mathrm{a}$\,$\parallel$\,$c$ is limited by the orbital effect, the weaker orbital effect for $B_\mathrm{a}$\,$\parallel$\,$ab$ due to the mass anisotropy is partly compensated for by the spin-paramagnetic effect leading to a reduced anisotropy at low temperatures.

We would like to thank M. Imai for the experimental supports, and M. Takigawa for the useful discussion.
The magnetic torque measurements were performed at the National High Magnetic Field Laboratory, which is supported by National Science Foundation Cooperative Agreement No. DMR-0654118, the State of Florida, and the U.S. Department of Energy.
A. K. was supported by NSF-DMR 0602859


\begin{thebibliography}{9}
\bibitem{Sample}
All preparative procedures were performed in a glove box filled with purified Ar gas.
Until the sample was mounted on the cantilever, it had been kept in argon atmosphere or in a vacuum desiccator.
\bibitem{TorqueSignal}
A torque signal of 1 in the units of the vertical axis in Figs.~\ref{fig2} and~\ref{fig4} corresponds to a 4\,\% change in the piezoresistance.
Since the linear response region of the cantilever is limited, typically below a few percentage of the resistance change, the shapes of hysteresis loops with large torque, e.g., one at $\theta$\,=\,60\,$^\circ$ (0.3\,K) in Fig.~\ref{fig4}(a), may be distorted.
\bibitem{Bc2}
The field defined here is, indeed, the irreversibility field, 
but we can safely assume that it is practically identical to $B_\mathrm{c2}$ for this compound, which assumption is justified by the fact that thus determined $B_\mathrm{c2}$ values are in good agreement with those determined from heat capacity and NMR measurements (see Fig.~\ref{fig3}).
\bibitem{Tc}
The WHH fit with $T_\mathrm{c}$ being a parameter yields $T_\mathrm{c}$\,=\,15.7 and 15.2\,K for $B$\,$\parallel$\,$ab$ and $B$\,$\parallel$\,$c$, respectively.
In this study, the average value of 15.5\,K is used as $T_\mathrm{c}$ of LiFeAs.
\end{thebibliography}
\end{document}